\documentstyle[12pt,lnfprepCMS,epsfig,amsmath]{article}
\newcommand{\be}{\begin{equation}}
\newcommand{\ee}{\end{equation}}
\def\simge{\mathrel{%
      \rlap{\raise 0.511ex \hbox{$>$}}{\lower 0.511ex \hbox{$\sim$}}}}
\def\simle{\mathrel{
      \rlap{\raise 0.511ex \hbox{$<$}}{\lower 0.511ex \hbox{$\sim$}}}}
\newcommand{\Header}{
     \begin{tabular}{rl}
     \hspace{-.3cm}
     \includegraphics{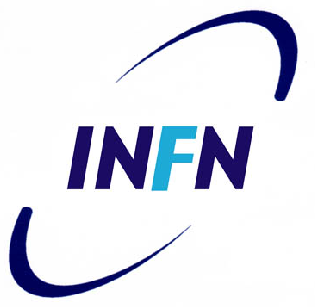} 
            &
       \renewcommand{\arraystretch}{0.5}
       \begin{tabular}{r}
         {\hspace{1cm}~\LARGE\sffamily LABORATORI~ NAZIONALI~ DI~ FRASCATI}\\
         \\
         {\Large\sffamily SIS-Pubblicazioni}\\
       \end{tabular}
       \renewcommand{\arraystretch}{1}
     \end{tabular}
  \vskip 1cm
  \begin{flushright}
  \renewcommand{\arraystretch}{0.5}
    \begin{tabular}{r}
      {\underline{LNF--08/27(P)}}\\    
      {\small November 29, 2008} \\      
      \\
    \end{tabular}
  \end{flushright}
  \renewcommand{\arraystretch}{1}
  \vskip 1 cm
  }

\begin{document}
\begin{titlepage}
\title{ 
  \Header
  {\large \bf SENSITIVITY AND ENVIRONMENTAL RESPONSE OF THE
CMS RPC GAS GAIN MONITORING SYSTEM}
}
\author{
L.~Benussi$^1$,
S.~Bianco$^1$,
S.~Colafranceschi$^1,2,3$,
F.~L.~Fabbri$^1$,
M.~Giardoni$^1$
\\
B.~Ortenzi$^1$,
A.~Paolozzi,$^1,2$
L.~Passamonti$^1$,
D.~Pierluigi$^1$
\\
B.~Ponzio$^1$,
A.~Russo$^1$,
A.~Colaleo$^4$,
F.~Loddo$^4$,
M.~Maggi$^4$
\\
A.~Ranieri$^4$,
M.~Abbrescia$^4,5$,
G.~Iaselli$^4,5$,
B.~Marangelli$^4,5$,
S.~Natali$^4,5$
\\
S.~Nuzzo$^4,5$,   
G.Pugliese$^4,5$,
F.~Romano$^4,5$,
G.~Roselli$^4,5$,
R.~Trentadue$^4,5$
\\
S.~Tupputi$^4,5$,
R.~Guida$^3$,
G.~Polese$^3,6$,
A.~Sharma$^3$,
A.~Cimmino$^7,8$
\\
D.~Lomidze$^8$,
D.~Paolucci$^8$,
P.~Piccolo$^8$,
P.~Baesso$^9$,
M.~Necchi$^9$
\\
D.~Pagano$^9$,
S.~P.~Ratti$^9$,
P.~Vitulo$^9$,
C.~Viviani$^9$\\
{\it ${}^{1)}$ INFN Laboratori Nazionali di Frascati, Via E. Fermi 40,
I-00044 Frascati, Italy.} \\
{\it ${}^{2)}$ Universit\`a degli Studi di Roma "La Sapienza",
Piazzale A. Moro.} \\
{\it ${}^{3)}$ CERN CH-1211 Gen\'eve 23 F-01631 Switzerland.} \\
{\it ${}^{4)}$ INFN Sezione di Bari, Via Amendola, 173I-70126 Bari,
Italy.} \\
{\it ${}^{5)}$ Dipartimento Interateneo di Fisica, Via Amendola,
173I-70126 Bari, Italy.} \\
{\it ${}^{6)}$  Lappeenranta University of Technology, P.O. Box 20
FI-538 1 Lappeenranta, Finland.} \\
{\it ${}^{7)}$ INFN Sezione di Napoli, Complesso Universitario di
Monte Sant'Angelo, edificio 6, 80126 Napoli, Italy.} \\
{\it ${}^{8)}$ Universit\`a di Napoli Federico II, Complesso
Universitario di Monte Sant'Angelo, edificio 6, 80126 Napoli,
Italy.}\\
{\it ${}^{9)}$ INFN Sezione di Pavia, Via Bassi 6, 27100 Pavia, Italy and Universit\`a degli studi di Pavia, Via Bassi 6, 27100 
Pavia, Italy.} 
}
\maketitle
\baselineskip=14pt

\begin{abstract}
Results from the gas gain monitoring
 (GGM) system for the muon detector using RPC in the CMS experiment at
the LHC is presented.
 The system is designed to provide
fast and accurate determination of any shift in the working point of the chambers due to gas
mixture changes.
\end{abstract}

\vspace*{\stretch{2}}
\begin{flushleft}
\end{flushleft}
\begin{center}

{\it{Presented by S.~Colafranceschi at the IEEE 08 - 
23 October 2008, Dresden, Germany}}
\end{center}
\end{titlepage}
\pagestyle{plain}
\setcounter{page}1
\baselineskip=17pt

\section{Introduction}
 Resistive Plate Chambers (RPC) detectors are widely used in HEP experiments for muon
detection and triggering at high-energy, high-luminosity hadron colliders, in
astroparticle physics experiments for the detection of extended air showers, as well as in
medical and imaging applications. At the LHC, muon systems of the CMS experiment
rely on Drift Tubes (DT), Cathode Strip Chambers (CSC) and RPCs for their muon trigger system, with a total gas volume
of about 50 m$^3$.  Utmost attention has to be paid to the possible
presence of gas contaminants which degrade the chamber performance.
The gas gain monitoring (GGM) system
monitors the gas quality online and is based on small RPC
detectors. The working point - gain and efficiency - is continuously monitored
along with environmental parameters, such as
temperature, pressure and humidity, which are important for
 the operation of the muon detector system.
 Design parameters, construction, prototyping and preliminary commissioning
results of the CMS RPC Gas Gain Monitoring (GGM) system have been  presented previously
\cite{Abbrescia:2007mu},\cite{Benussi08}. In this paper, results on the response of the GGM 
detectors to
environmental changes are presented.
 \par
 The CMS
RPCs are bakelite-based double-gap RPC with strip readout (for construction
details see
\cite{:2008zzk} and reference therein) operated with 96.2\% C$_2$H$_2$F$_4$ -
3.5\%
Iso-C$_4$H$_{10}$ - 0.3\% SF$_6$ gas mixture humidified at about 40\%. The large
volume of the
whole CMS RPC system and the cost of gas used make mandatory the operation of
RPC in a
closed-loop gas system (for a complete description see \cite{gassystem}), in
which the gas fluxing the gaps is reused after being purified by a set of
filters\cite{purifiers}.

The operation of the CMS RPC system is strictly correlated to the ratio between
the gas mixture components, and to the presence of pollution due to contaminants
that can be  be produced inside the gaps during discharges (i.e. HF produced by
SF$_6$ or C$_2$H$_2$F$_4$
molecular break-up and further fluorine recombination), accumulated in the
closed-loop or by
pollution that can be present in the gas piping system (tubes, valves, filters,
bubblers, etc.)
and flushed into the gaps by the gas flow. The monitoring of the presence of these
contaminants, as
well as the gas mixture stability,  is therefore mandatory to avoid RPC damage
and to ensure their correct functionality.

A monitoring system of the RPC working point due to changes of gas composition  
and pollution must provide a faster and sensitive response than the CMS RPC
system itself in order
to avoid irreversible damage of the whole system. Such a Gas Gain Monitoring
system  monitors
efficiency and signal charge continuously by means of a  cosmic ray
telescope based
on RPC detectors. In the following will be briefly described the final setup of
the GGM system, and
the first results obtained during  its commissioning  at the ISR test area
(CERN).

\section{The Gas Gain Monitoring System}
The GGM system  is composed by the same type of RPC used in the CMS detector but
of smaller size (2mm Bakelite gaps, 50$\times$50 cm$^2$).
Twelve gaps are arranged in a stack located in the CMS gas
area (SGX5 building) in the surface, close to CMS assembly hall (LHC-P5). The
choice to install the
system in the surface instead of underground allows one to profit from maximum
cosmic muon rates. In order to ensure a fast response to working point shifts 
with a precision
of 1\%, $10^4$ events are are required, corresponding to
about 30 minutes exposure time on surface, to be compared with a 100-fold lesser
rate underground. The trigger is
provided by four out of twelve gaps of the stack, while the remaining eight gaps
are used to monitor the working point stability.
\par
The eight gaps are arranged in three sub-system: one sub-system (two gaps)
is fluxed with the fresh CMS mixture and its output sent to vent. The second 
sub-system (three gaps) is fluxed with CMS gas coming from the closed-loop  
gas system and extracted before the gas purifiers, while the third sub-system
(three gaps) is operated with CMS gas extracted from the closed-loop extracted
after the gas filters.
The basic idea is to compare the operation of the three sub-systems and, if some
changes are observed, to send a warning to the experiment. In this way, the gas
going to and coming from the CMS RPC detector is monitored by using the two gaps
fluxed with the fresh mixture as reference gaps. This setup will ensure that
pressure, temperature and humidity changes affecting the gaps behavior do cancel
out by comparing the response of the three sub-system operating in the same
ambient condition. 
\par
The monitoring is performed by measuring the charge distributions of each
chamber. The eight gaps are operated at different high voltages, fixed for
each chamber, in order to monitor the total range of operating modes of the   
gaps. The operation mode of the RPC changes as a function of the voltage
applied. A fraction of the eight gaps will work in pure avalanche mode, while
the remaining will be operated in avalanche+streamer mode. Comparison of  signal
charge distributions and the ratio of the avalanche to streamer components of
the ADC provides a monitoring of the stability of working point for changes due 
to gas mixture variations.
\par
\begin{figure}[H]
   \begin{center}
    \resizebox{0.37\textwidth}{!}{\includegraphics{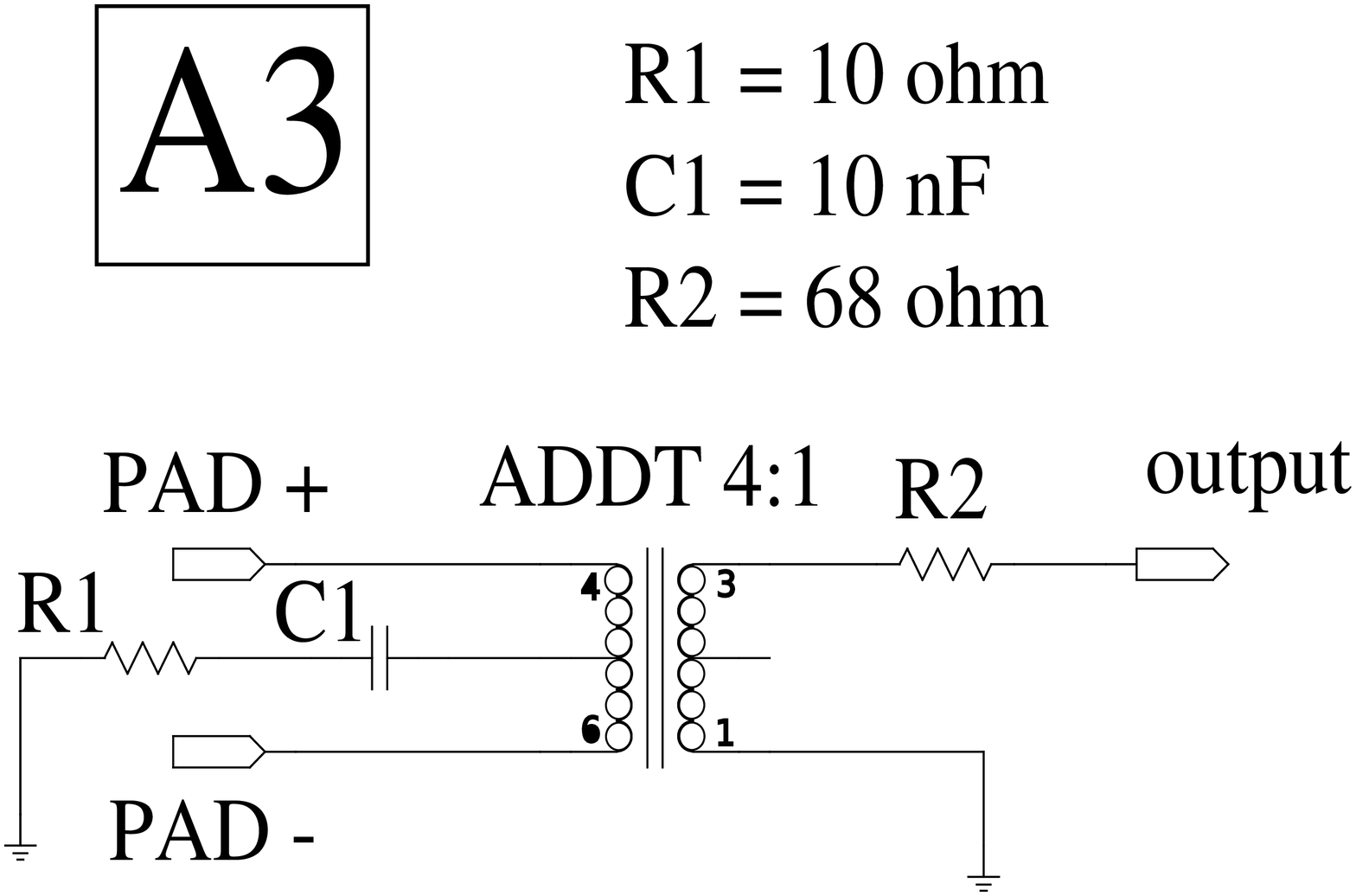}}
    \caption{The electric scheme of the read-out circuit providing the algebraic
sum of the two      pad signal (PAD + and PAD -).}
    \label{fig3}
    \end{center}
  \end{figure}

Details on the construction of GGM can be found in \cite{Benussi08}.
Each chamber of the GGM system consists of a single gap with double sided pad
read-out: two copper pads are glued on the two opposite external side of the
gap.  The signal
is read-out by a transformer based circuit A3 (Fig.\ref{fig3}). The
circuit allows to algebraically subtract the two signal, which have opposite polarities,
and to obtain an output signal with subtraction of the coherent noise, with an  
improvement by about a factor 4 of the signal to noise ratio.
The output signals from circuit A3 are sent to a CAEN V965 ADC \cite{caen} for  
charge analysis.

\begin{figure}[H]
  \begin{center}
   \resizebox{0.44\textwidth}{!}{\includegraphics{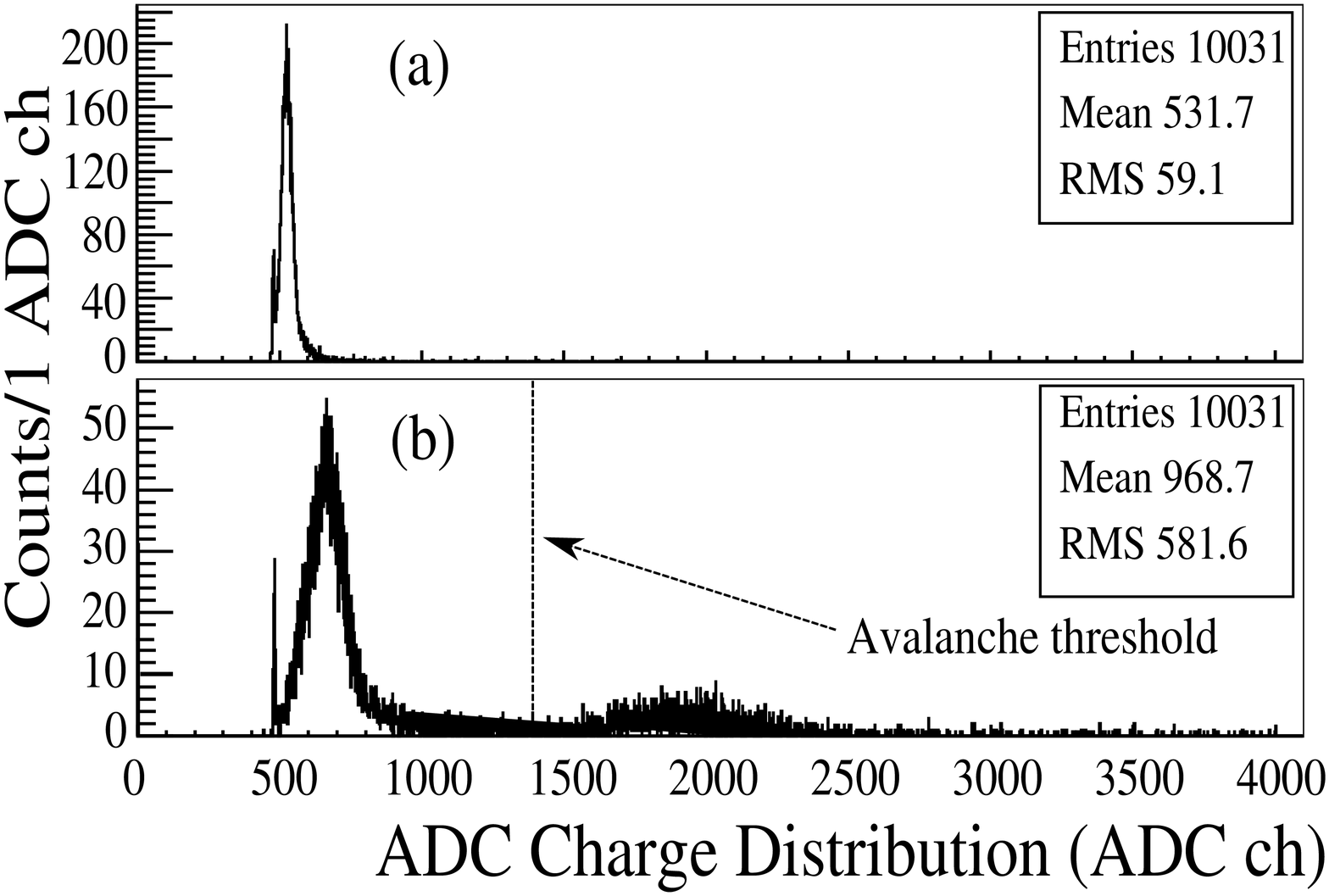}}
    \caption{Typical ADC charge distributions of one GGM chamber at two
operating voltages. Distribution (a) correspond to HV$_{eff}$ = 9.9kV while
distribution (b) to
      HV$_{eff}$=10.7kV. In (b) is clearly visible the streamer peak around 1900
ADC channels.
      The events on the left of the vertical line (1450 ADC channels in this
case) are assumed to be pure avalanche events.}
    \label{fig6}
    \end{center}
  \end{figure}

A typical ADC distribution of a GGM gap is shown in fig.\ref{fig6} for two
different effective
operating voltage,  defined as the high voltage set on the HV power supply
corrected for the local
atmospheric pressure and temperature. Fig.\ref{fig6} a) corresponding to
HV$_{eff}$=9.9kV shows a clean avalanche peak well separated from the pedestal.
Fig.\ref{fig6} b) shows the charge distribution
at HV$_{eff}$=10.7kV with two signal regions corresponding to the avalanche and
to avalanche+streamer mode.

  \begin{figure}[H]
   \begin{center}
    \resizebox{0.44\textwidth}{!}{\includegraphics{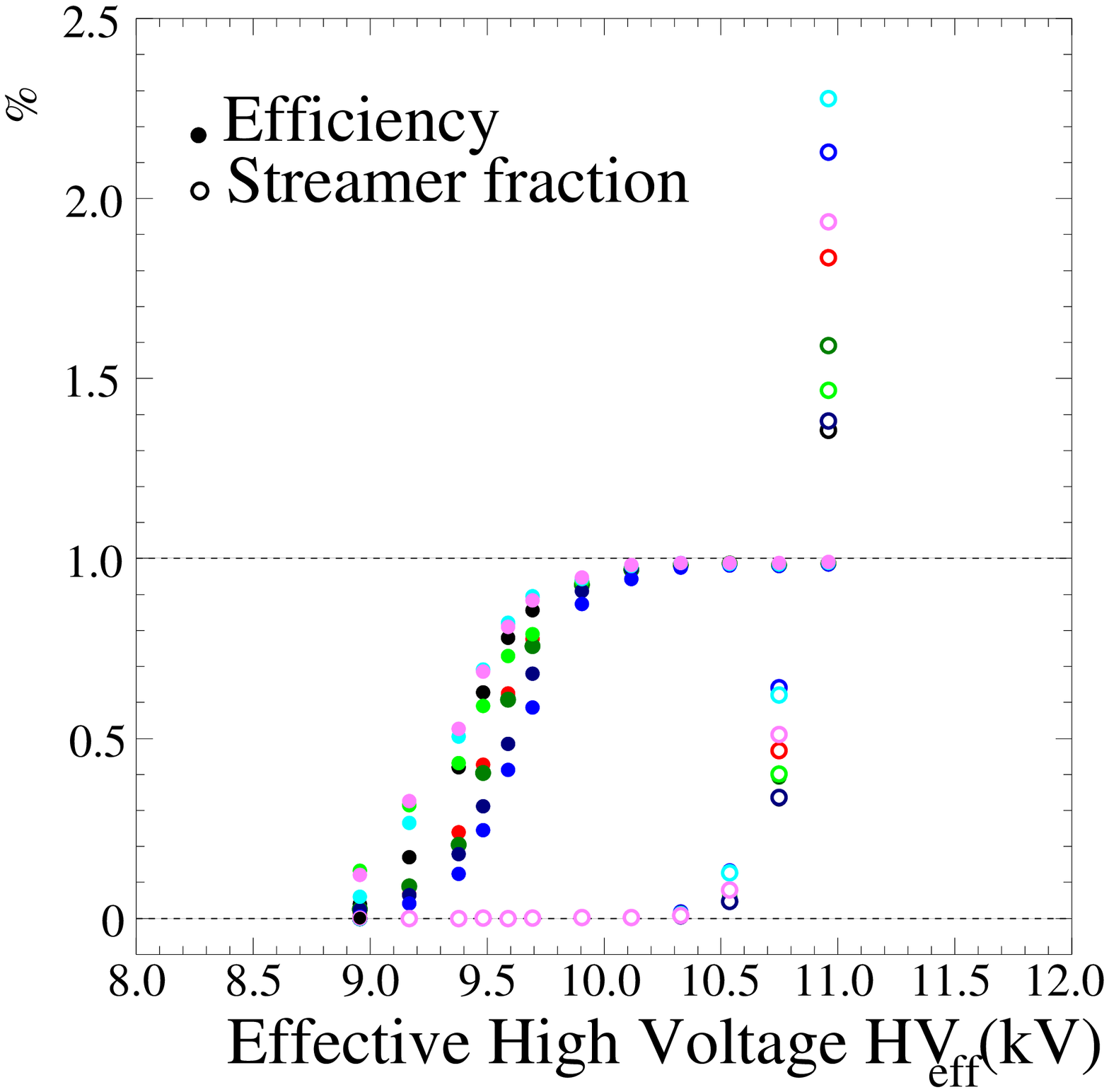}}
    \caption{Efficiency plot (full dots) of GGM chambers as a function of
HV$_{eff}$. The efficiency is
      defined as the ratio between the number of ADC entries above
3$\sigma_{ped}$ and the number
      of acquired triggers. Open dot plots correspond to the streamer fraction 
of the chamber
      signal as a function of HV$_{eff}$.}
    \label{fig7}
    \end{center}
  \end{figure}
Fig.\ref{fig7} shows the GGMS single gap efficiency (full dots), and the ratio
between the
avalanche and the streamer component (open circles), as a function of
the effective high voltage. Each point corresponds to a total of 10000 entries
in the full ADC spectrum. The efficiency is defined as the ratio between the
number of triggers divided by the
number of events above 3$\sigma_{ped}$ over ADC pedestal, where $\sigma_{ped}$
is the pedestal width.
The avalanche to streamer ratio is defined
by counting the number of entries in the avalanche (below the ADC threshold
(fig.\ref{fig6} b)
and above the pedestal region) and dividing it by the number of streamer events above the
avalanche threshold. Both
efficiency and avalanche plateau are in good agreement with previous results
\cite{Abbrescia:2005yr}.
      
\begin{figure}[H]
\begin{center}
    \resizebox{0.44\textwidth}{!}{\includegraphics{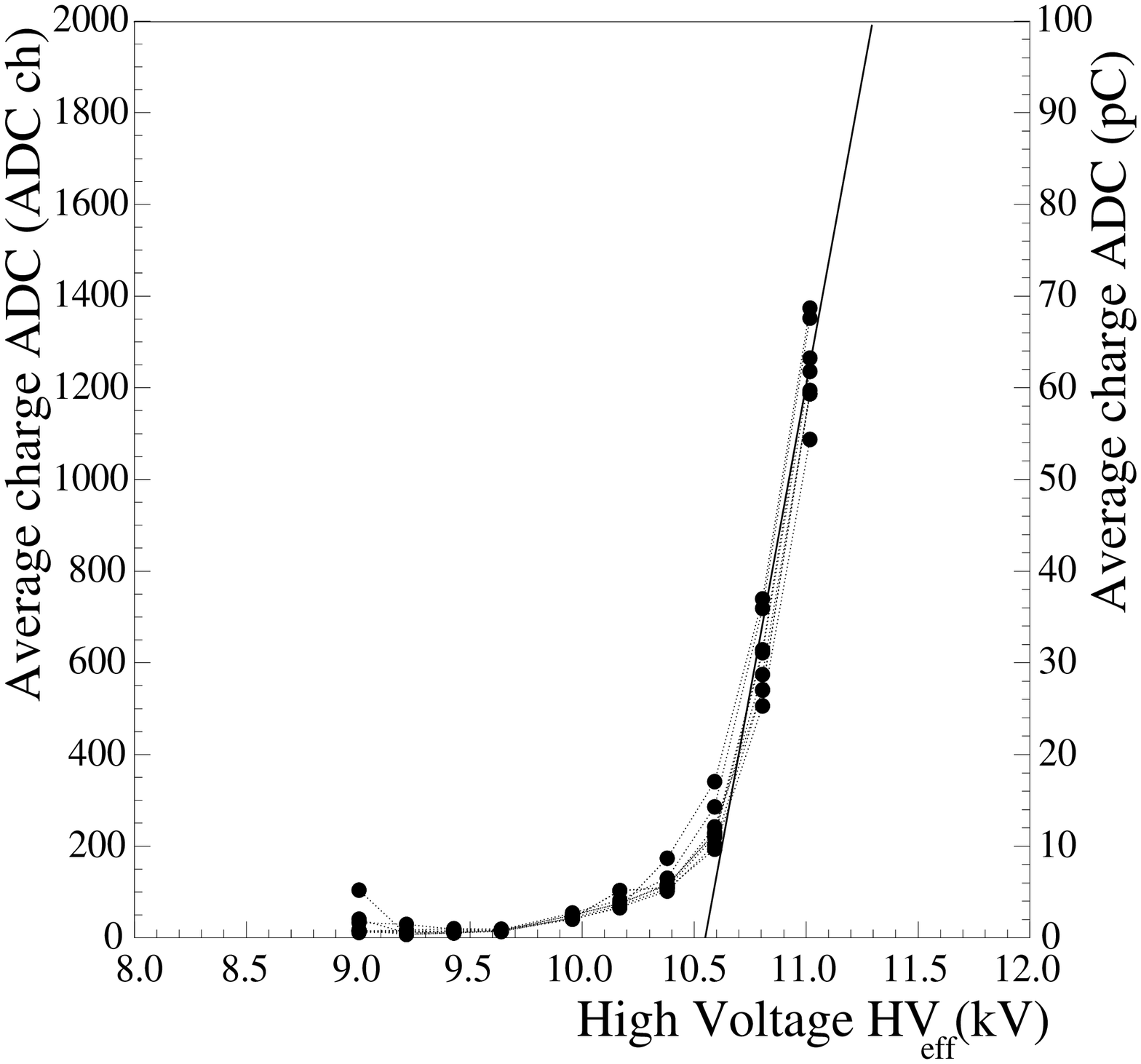}}
    \caption{Avarege avalanche charge of the eight monitor chamber signal as a
function of HV$_{eff}$.
     The slope is about 25 ADC ch/10 or 1.2pC/10V. Each point corresponds to
10000 triggers.}
    \label{fig8}
\end{center}
  \end{figure}

In order to determine the sensitivity of GGM gaps to working point shifts, the
avalanche to streamer transition was studied by two methods, the charge method 
and the efficiency method. In the charge method, the mean value of the ADC
charge distribution in the whole ADC range is studied as a function of
HV$_{eff}$ (fig.\ref{fig8}). Each point corresponds to 10000 events in the
whole ADC spectrum. In the plot three working point regions are identified
\begin{enumerate}
   \item inefficiency (HV$_{eff}<$ 9.7 kV);
   \item avalanche (9.7 kV $<$  HV$_{eff}<$ 10.6 kV;
   \item avalanche+streamer mode (HV$_{eff}>$ 10.6 kV).
\end{enumerate}
The best sensitivity to working point shifts is achieved in the
avalanche+streamer region, estimated to be about 25~ADC~ch/10~V or 1.2pC/10V.
      
\begin{figure}[H]
  \begin{center}
    \resizebox{0.44\textwidth}{!}{\includegraphics{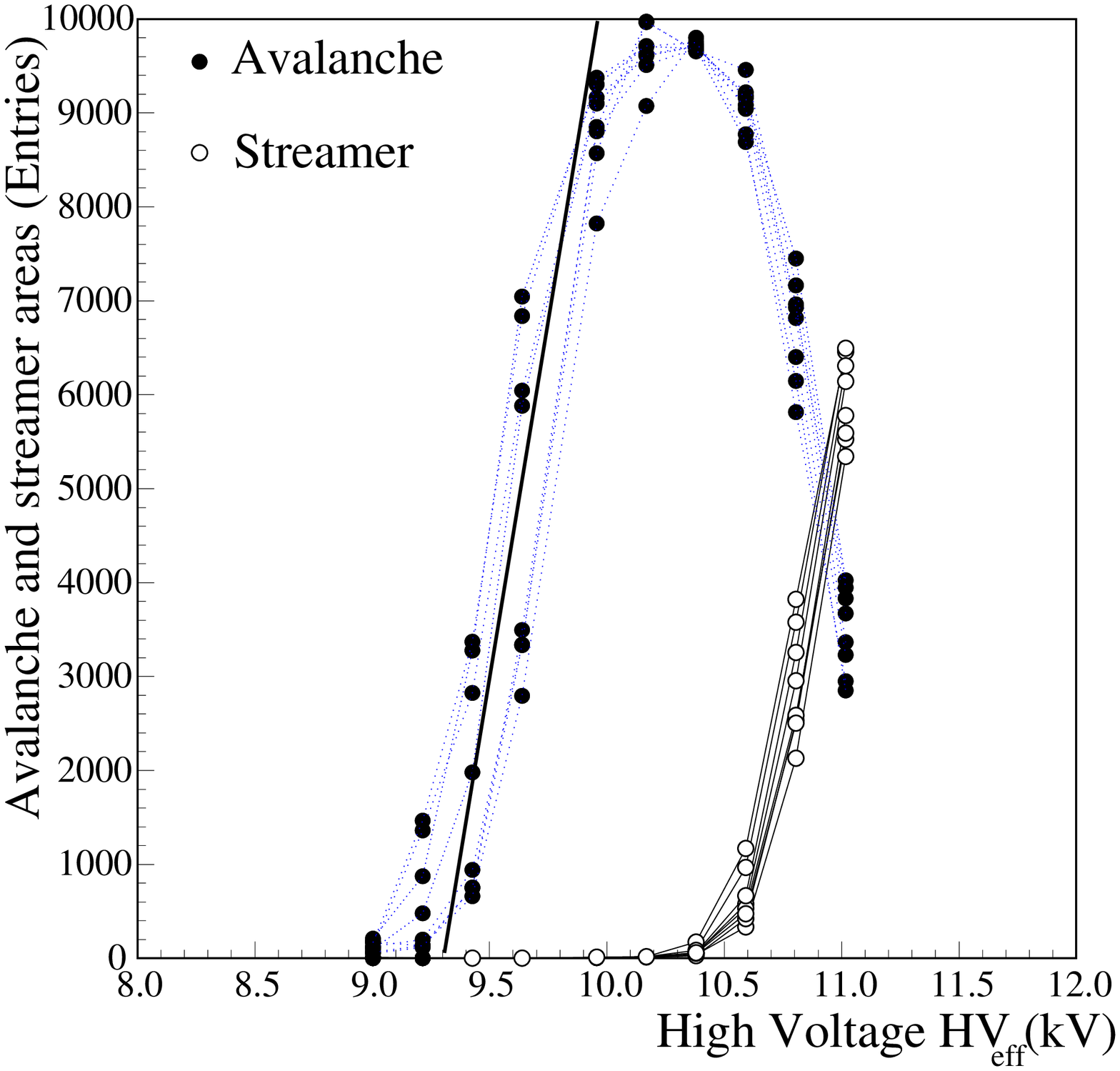}}
    \caption{Streamer and avalanche yields as a function of HV$_{eff}$. Each
point corresponds to
      10000 collected triggers. The solid line has a slope of approximately 130
events/10 V corresponding to a sensitivity of 1.3\%/10V.}
    \label{fig9}   
\end{center}
  \end{figure}

In the efficiency method, the ADC avalanche event yield is studied
 as a function of HV$_{eff}$ (\ref{fig9}). The avalanche signal
 increases by increasing the HV applied to the gap, until it reaches a maximum
value after which
the streamer component starts to increase. The 9.0kV-10.0kV shows a sensitivity
to work point changes of approximately  1.3/\%/10V.

\section{Response of GGM to environmental effects}
The workpoint of GGM is affected by environment. However, all
environmental effects cancel out thanks to redundancy
of the system. Each
environmental effect not connected with a modification of gas mixture will be cancelled out
by a comparison between different RPC
chamber flown with the same gas, which are affected by the same environmental parameters.
\par
\begin{figure}[H]
  \begin{center}
    \resizebox{0.50\textwidth}{!}{\includegraphics{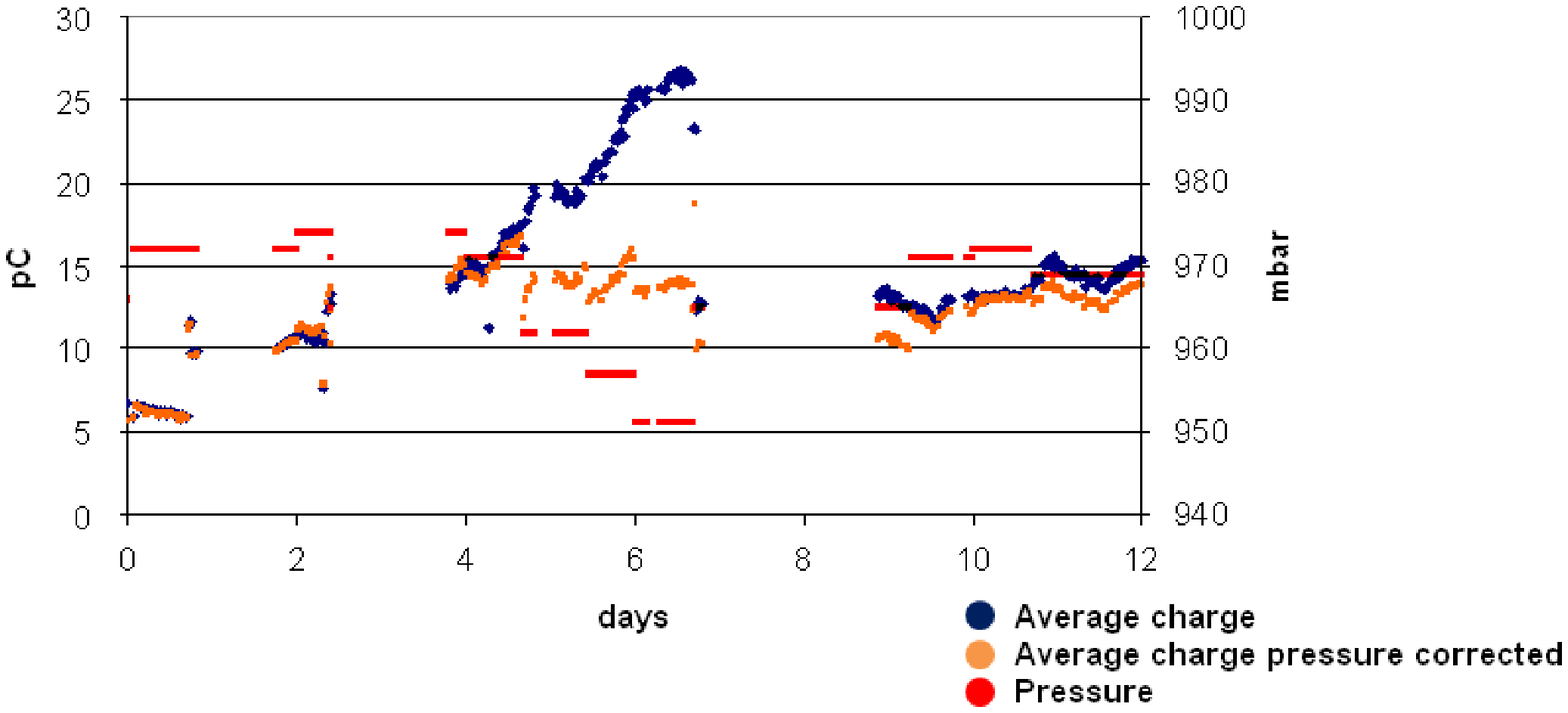}}
    \caption{Average charge and pressure-corrected charge}   
    \label{picchetto}
    \end{center}
  \end{figure}
\begin{figure}[H]
  \begin{center}
    \resizebox{0.50\textwidth}{!}{\includegraphics{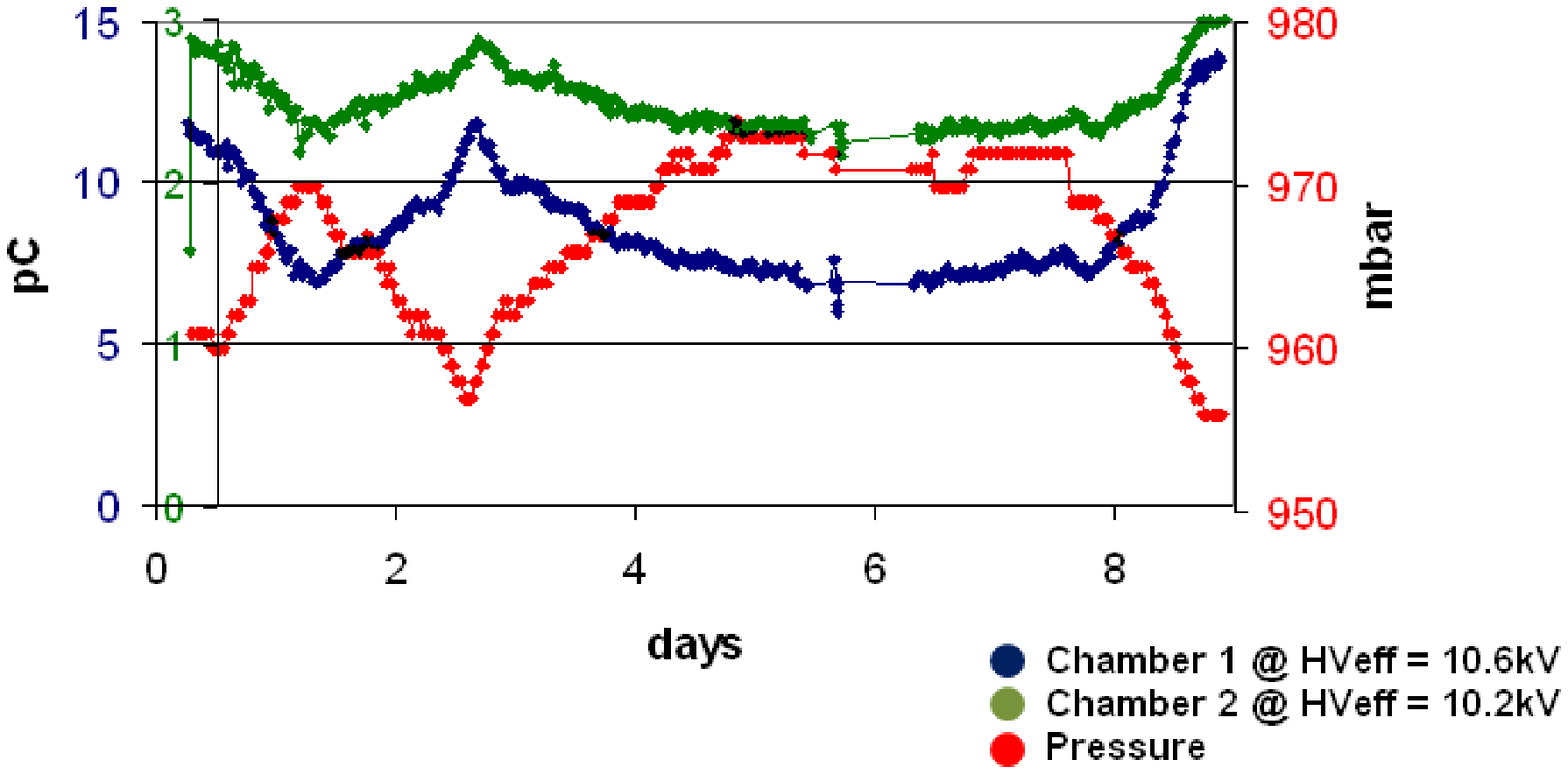}}
    \caption{Average charge of two chambers at different voltages as influenced by pressure}
    \label{ch1ch2pr}
    \end{center}
  \end{figure}

An example of such cancellation is shown in Fig.~\ref{picchetto}, where the average charge
distribution (black dots) is plotted across a changeover of gas bottles.
 Data show a sudden increase in the
average charge distribution which may interpreted as a shift of working point due to changes
in gas mixture composition. By weighing the average charge with a correction factor linearly
depending on atmosferic pressure, however, no significant increase   is left in the
distribution of corrected average charge (green dots) which may signal
an anomalous shift due to gas mixture.
The cancellation algorithm is applied by correcting variables withing gaps belonging to
the same subdetector. Fig.~\ref{ch1ch2pr} shows the average charge for two chambers
  working in different regimes at different voltages. The average charge of both chambers is
  completely correlated, and very well correlated to the atmosferic pressure variations.
\par


\section{Conclusions}
Results from the  Gas Gain Monitoring System for the CMS RPC Detector have been reported on.
The purpose of GGM is to monitor any shift of the
working point of the CMS RPC detector. The GGM is being commissioned at CERN  and is planned
to start  operation by the end of 2008. Preliminary
results show good sensitivity to working point changes.
 The system redundancy allows for effectively cancelling out the environmental effects.
Further tests are in progress to
determine the sensitivity to gas variations.
%

%
\end{document}